\documentclass[%
reprint,
 amsmath,amssymb, 
 longbibliography, 
 aps,prl]{revtex4-1}

\usepackage{float}
\usepackage{graphicx}
\usepackage{dcolumn}
\usepackage{bm}
\usepackage[flushleft]{threeparttable}
\usepackage{url}
\usepackage{amsmath}
\usepackage{amssymb}
\usepackage{braket}
\usepackage{multirow}
\usepackage{array}
\usepackage{url}
\newcolumntype{P}[1]{>{\centering\arraybackslash}p{#1}}
\usepackage{xcolor}
\usepackage{framed}
\usepackage{lipsum}
\colorlet{shadecolor}{blue!60}
\usepackage{soul}
\usepackage{hyperref}
\usepackage{upgreek}
\usepackage{longtable}
\usepackage{ulem}
\makeatletter

\begin{document}

\preprint{APS/123-QED}

\title{Competing phases of HfO$_2$ from unstable flat phonon bands of an unconventional high-symmetry structure}

\author{Yubo Qi$^{1}$, and Karin M. Rabe$^2$}

\affiliation{%
$^1$ Department of Physics, University of Alabama at Birmingham, Birmingham, Alabama 35233, USA \\
$^2$Department of Physics $\&$ Astronomy, Rutgers University, 
Piscataway, New Jersey 08854, United States 
}%

\date{\today}

\begin{abstract}

We carry out first-principles calculations to demonstrate that the complex energy landscape and competing phases of HfO$_2$ can be understood from the four unstable flat phonon bands of an unconventional high-symmetry structure of HfO$_2$ with the space group $Cmma$.
We consider structures generated from the $Cmma$ reference structure by all possible combinations of the zone center and zone boundary modes belonging to the unstable flat phonon branches. 
We find 12 distinct locally-stable structures, of which 5 correspond to well-known phases.
We also show that 8 of these 12 structures can be described as period-2 superlattices of the ferroelectric $Pca2_1$ (oIII), ferroelectric $Pnm2_1$ (oIV), monoclinic $P2_1/c$ (m) and distorted monoclinic $P2_1/c$ (dm) structures.
We demonstrate how the unstable flat phonon bands can explain the atomically thin grain boundaries in the various types of superlattices.
Finally, we point out that arbitrary-period HfO$_2$ superlattices derived from the 6 different types of period-2 superlattices are expected to form based on the flatness of the unstable phonon branches.
The organizing principle provided by this work deepens our understanding of the underlying physics in the phase stability of HfO$_2$ and provides guidance for functional phase stabilization.  

\end{abstract}

\pacs{Valid PACS appear here}

\maketitle

HfO$_2$-based ferroelectrics are highly promising for technological applications due to their robust polarization in thin films and outstanding compatibility with silicon~\cite{Schroeder22p653,Park18p795,Park15p1811,Muller12p4318,Cheema22p2100499,Cheema22p65}.
The discovery of ferroelectricity in HfO$_2$ stimulated intense research activity, revealing that HfO$_2$ has a rich energy landscape with many competing phases~\cite{Liu19p054404,Huan14p064111,Behara22p054403,Ma23p096801}. 
These include the fluorite cubic (denoted as c, space group $Fm\overline{3}m$)~\cite{Wang92p5397} structure, tetragonal (denoted t, space group $P4_2/nmc$) phase, monoclinic (denoted m, space group $P2_1/c$) phase~\cite{Wang92p5397}, ferroelectric orthorhombic (denoted oIII, space group $Pca2_1$) phase~\cite{Boscke11p102903,Boscke11p112904,Sang15p162905,Xu21p826,Qi20p214108,Mueller12p2412,Pevsic16p4601,Muller11p114113,Olsen12p082905}, second ferroelectric (denoted oIV, space group $Pnm2_1$) phase~\cite{Qi20p257603,Huan14p064111}, and an anti-polar orthorhombic (denoted o-AP, space group $Pbca$) phase~\cite{Lowther99p14485,Jayaraman93p9205,Ohtaka01p1369}. 

In addition, studies of ferroelectric HfO$_2$ show that the domain walls between polar variants are atomically thin, and domain widths can be as small as a unit cell~\cite{Lee20p1343,Marquardt23p3251,Choe21p8,Ding20p556}.
In fact, it has been observed that the grain boundaries in the m phase, and between two competing phases, such as the m and the oIII phase, can also be atomically thin~\cite{Aoki05p450,Chirumamilla19p7241,Grimley18p1701258}.
These localized grain boundaries can significantly influence the material's overall properties, such as mechanical strength and electrical properties, contributing to its unique and complex behavior in various applications~\cite{Lee20p1343,Degoli17p2404378,Shubhakar16p204,Petzold19p1900484,Grimley18p1701258}.
The ultra-thin domain walls and domains in the ferroelectric phase have been attributed to a non-soft flat phonon band at about 200 cm$^{-1}$ in the cubic fluorite structure, which persists in the oIII phase~\cite{Lee20p1343}. 

In this work, we present first-principles calculations of the structure and phonon dispersion of the $Cmma$ structure of HfO$_2$, identified as a reference structure through consideration of polarization switching paths. 
From the flat unstable bands in this phonon dispersion, we identify 4 unstable lattice distortions at the zone center and 4 at the zone boundary, and from these generate $2^8=256$ candidate low-energy structures, which relax to 12 locally stable structures, of which 5 correspond to known phases. 
We show that 8 of the 12 phases can be recognized as period-2 unit-cell-scale superlattices composed of the known phases with atomically thin grain boundaries, and that this is a natural consequence of the flatness of the unstable phonon bands. 
Further, the flatness of the bands suggests that longer period superlattice structures will also be present in the energy landscape, which is confirmed by our first-principles calculation.
The implications of this for understanding and predicting the physical functionaility of HfO$_2$ are discussed.

\begin{figure}[hpbt]
\includegraphics[width=7.5cm]{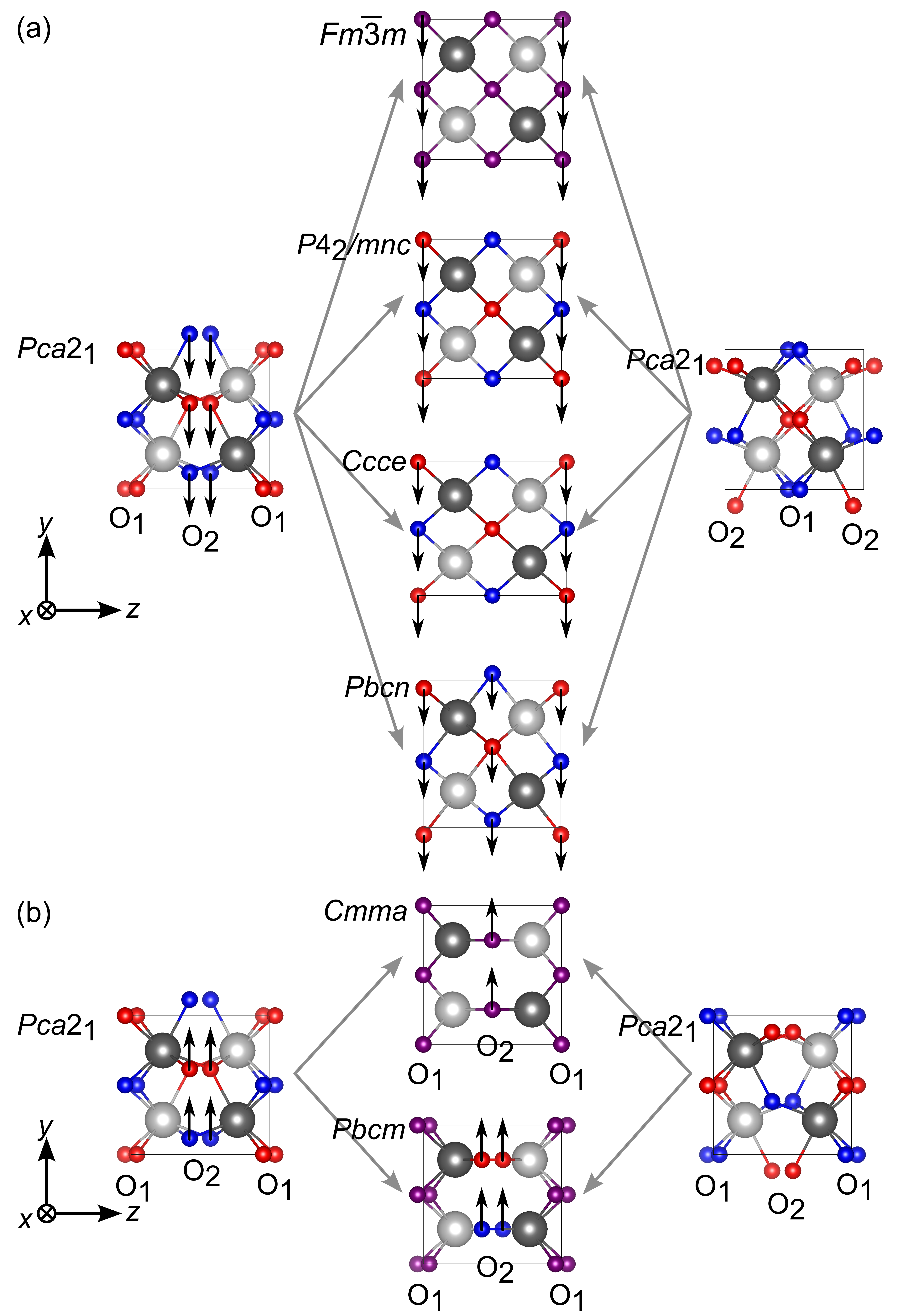}
\caption{(a) First and (b) second categories of polarization switching paths and reference structures used for lattice mode analysis along the two paths.
The dark and light grey spheres represent closer and farther Hf atoms viewing along the view ($x$) direction.  
Similarly, the red and blue colors indicate the oxygen atoms that are outward and inward displaced along the view direction.
The purple color indicates the oxygen atoms with negligible displacements along the view direction.
The arrows indicate the direction of oxygen movement during the polarization switching. 
O1 and O2 are labels of the two Wyckoff positions occupied by oxygen in the $Pca2_1$ structures, 
and the large displacements of O2 are responsible for the polarization.
}
\label{f1}
\end{figure}

For HfO$_2$, the high-temperature cubic fluorite phase has the highest symmetry of all known phases. 
As such, it seemed an obvious choice for a high-symmetry reference structure~\cite{Reyes14p140103,Qi20p214108,Lee20p1343,Zhou22peadd5953,Ma23p096801}, but questions quickly emerged.
The only lattice instability in the cubic fluorite structure is an $X_2^-$ phonon, which when frozen in gives the observed tetragonal $P4_2/nmc$ structure.
This $P4_2/nmc$ structure is locally stable, with no soft mode indicating the presence of another low energy structure. In fact, five additional lattice modes [1 polar and 4 antipolar modes, see Supplementary Materials (SM)~\cite{SM} Section 1]~\cite{Qi2021polarization} with higher-order couplings (in particular, trilinear coupling) are needed to describe the $Pca2_1$ ferroelectric phase~\cite{Reyes14p140103,Delodovici21p064405}. 
The cubic fluorite structure thus cannot be considered a natural high-symmetry reference structure for ferroelectricity in HfO$_2$.

\begin{figure*}
\includegraphics[width=17.5cm]{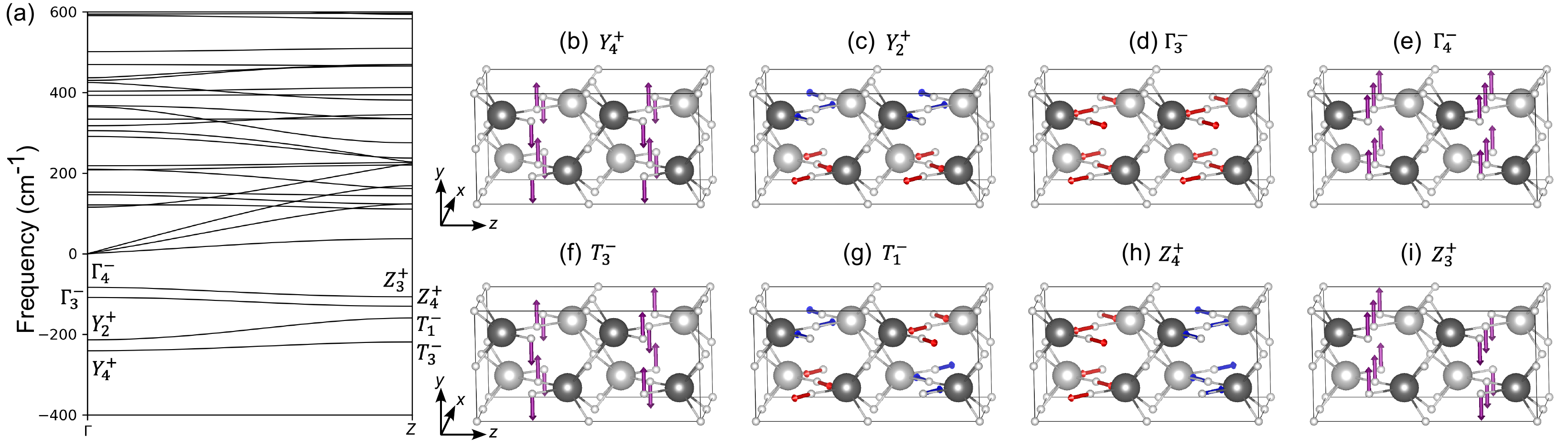}
\caption{(a) The phonon spectrum along $\Gamma-Z$ line of the conventional cell.
(b-i) Schematic illustrations of the phonon modes.
}
\label{f2}
\end{figure*}

A natural strategy for identifying a high-symmetry reference structure for a ferroelectric is to take the midpoint structure for a uniform switching path between two polarization variants.
Because of its complex structure, there are multiple plausible polarization switching paths in HfO$_2$, which have suggested additional candidate reference structures that have been discussed in the literature, including $Ccce$~\cite{Delodovici21p064405} and $Pbcn$~\cite{Raeliarijaona23p094109}.
The $Ccce$ structure is the $P4_2/nmc$ structure homogeneously strained to match the lattice of the $Pca2_1$ phase, with no additional atomic displacements, while the $Pbcn$ structure is generated from the $P4_2/nmc$ structure by an anti-polar mode characterized by alternating oxygen displacements along the $y$ direction.

In fact, recent studies have revealed that HfO$_2$ is a ``double-path'' ferroelectric, which means that HfO$_2$ has two competing categories of polarization switching paths for which the change in polarization is of opposite sign~\cite{Qi2022double,Wei22p154101,Choe21p8}. 
As shown in Fig.~\ref{f1} (a), in the first category of switching path, the oxygen atoms moving under an electric field stay within the original Hf tetrahedron. 
In such a category, the structure with the highest symmetry along the path is the cubic $Fm\overline{3}m$ structure.
In the second category of switching path [Fig.~\ref{f1} (b)], the oxygen atoms move across a plane containing Hf into an adjacent Hf tetrahedron.
The midpoint structure has a $Pbcm$ space group and is used as the reference structure in Ref.~\cite{Aramberri23p95,Choe21p8,Wei22p154101,Clima14p092906,Yang20p064012,Kingsland24p054102}.
The $Cmma$ structure is the highest-symmetry phase obtained from the midpoint $Pbcm$ structure through a simple symmetry-restoring distortion.
The symmetry group of the $Cmma$ structure is thus a supergroup of the symmetry group of the $Pbcm$ structure, as shown in Fig. S3.
Taking the $Cmma$ structure as the reference, the distortion in the $Pbcm$ phase corresponds to the $Y_2^+$ mode [Fig.~\ref{f2} (c)].

In the base-centered-orthorhombic $Cmma$ structure, half of the oxygens are at the center of the Hf tetrahedron, as in the cubic fluorite structure, and half are midway between Hf atoms in the $x-z$ plane. 
The $Cmma$ structure is a subgroup of the cubic fluorite phase, and there are 6 distinct variants of the $Cmma$ structure [see SM Section 3 for full structural information].
The calculated phonon spectrum of the $Cmma$ structure along the $\Gamma-Z$ line of the Brillouin zone of the conventional unit cell is shown in Fig.~\ref{f2} (a).
The $\Gamma-Z$ line of the Brillouin zone for the conventional cell combines the $\Gamma-Z$ and $Y-T$ lines 
of the primitive cell (see SM Section 4 for computational details and Section 5 for the full spectrum).
We see that along this line there are four unstable phonon branches, which we label according to the primitive cell, shown in Fig.~\ref{f2} (b-i).
The dispersions of these unstable phonon branches are 23 cm$^{-1}$, 22 cm$^{-1}$, 54 cm$^{-1}$, and 22 cm$^{-1}$ respectively, and thus they can be viewed as essentially flat.
With a flat phonon band, energy-lowering distortions in a unit-cell-thick layer can be made essentially independently from layer to layer.

\begin{figure*}
\includegraphics[width=17cm]{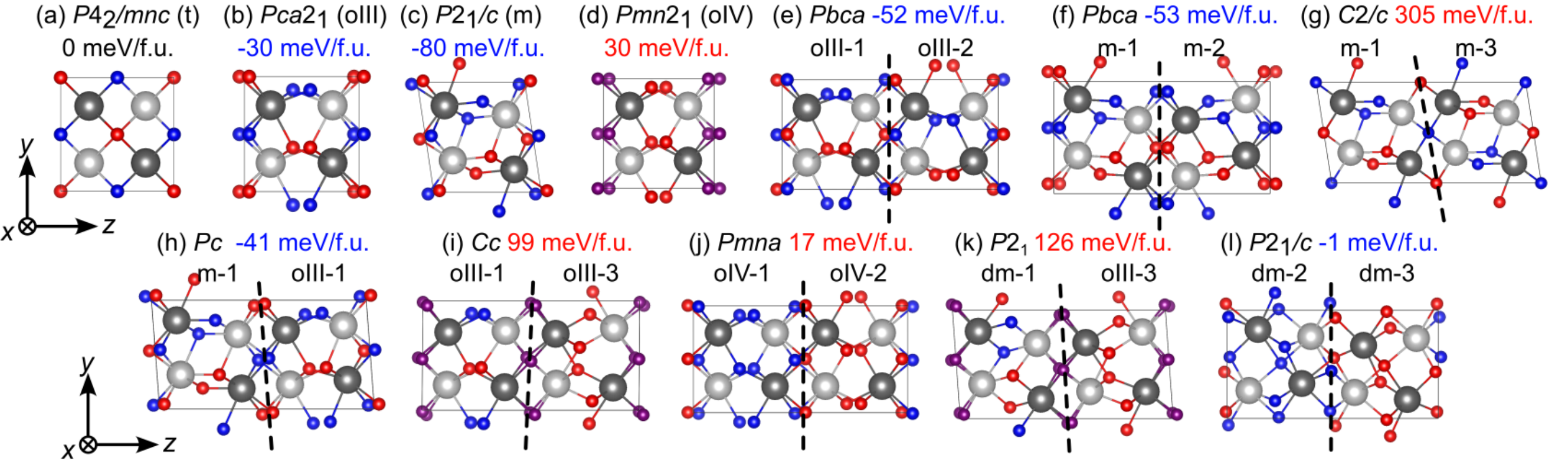}
\caption{(a-l) Superlattices/phases discovered from the analyses based on the unstable phonon modes of the $Cmma$ HfO$_2$.
The atomic color scheme follows the convention used in Fig.~\ref{f1}.
In labels like `oIII‑1' or `m‑1,' the numeric suffix denotes the variant index of that phase, and different variants are related by symmetry operations.
}
\label{f3}
\end{figure*}

The multiple unstable phonon bands in this state guide us to the identification of competing low-energy phases nearby in the energy landscape. 
To identify a local energy minimum, we start by ``activating'' (giving a nonzero amplitude to) one or more unstable phonon modes.
Consideration of the 4 unstable modes at $\Gamma$ and the 4 at $Z$, with each one either activated or not, leads to $2^8=256$ different distorted configurations.
We carry out first-principles calculations to relax each of these configurations, and find 12 distinct locally-stable structures and their variants correlated by symmetry operations (see SM Section 6 for detailed information about relaxed structures obtained by activation of various unstable modes).
The energies, space groups and mode amplitudes of these structures and their variants are shown in Tables S3 and S4.
5 of the 12 are well-known phases: t, oIII, m, oIV, and o-AP, as shown in Fig.~\ref{f3} (a-e), and the remaining 7 are shown in Fig.~\ref{f3} (f-l).
Their energies are also listed in Fig.~\ref{f3}, taking the energy of the t phase as the reference. 
All the 12 structures are locally stable, as confirmed by the calculated full phonon dispersion relations presented in SM section 7.
Moreover, as shown in Fig. S3, all 12 structures, except for the $P4_2/mnc$ t phase, are subgroups of the $Cmma$ structure, demonstrating its suitability as a reference. 
Analyzing the relaxation trajectory reveals that the $Cmma$ structure, under specific distortions, first relaxes into the cubic fluorite structure before evolving toward the $P4_2/mnc$ t phase.

\begin{figure}[hpbt]
\includegraphics[width=8.5cm]{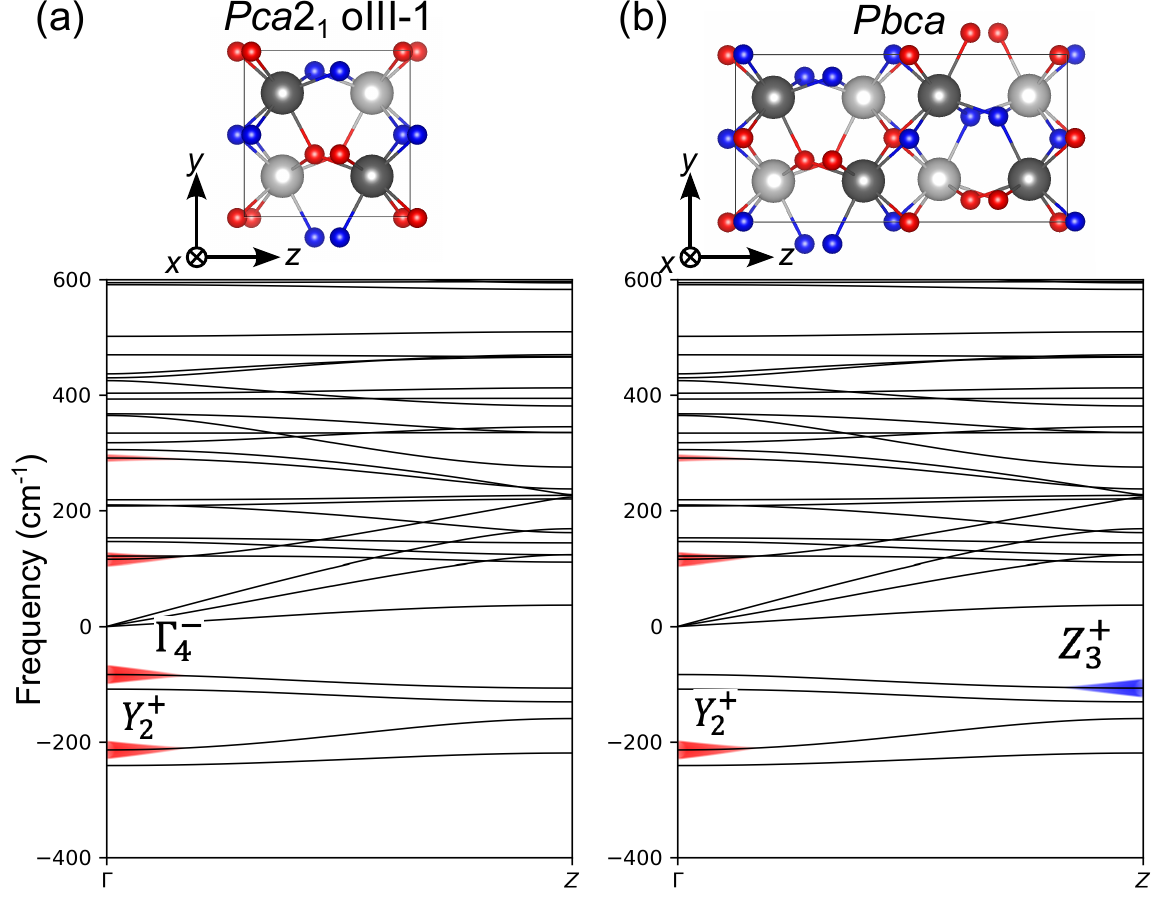}
\caption{Activation of the phonon modes for the (a) oIII $Pca2_1$ and  (b) $Pbca$ structures.
The activated zone center modes are highlighted with red and zone boundary modes are highlighted with blue.
}
\label{f4}
\end{figure}

The well-known o-AP phase of HfO$_2$ [Fig.~\ref{f3} (e)], whose energy is slightly lower than that of the oIII phase, can be viewed as a 180$^{\circ}$ unit-cell-scale domain structure of oIII with atomically-thin domain wall width, as has been discussed in Ref.~\cite{Lee20p1343}.
This can be naturally understood from the four unstable flat phonon branches.
As shown in Fig.~\ref{f4} (a), in the oIII phase, the $\Gamma_4^-$ and $Y_2^+$ modes are activated.
In the o-AP phase, the $\Gamma_4^-$ and $Z_3^+$ modes are activated [Fig.~\ref{f4} (b)].
The atomic distortion pattern of the zone boundary $Z_3^+$ mode is closely related to the $\Gamma_4^-$ mode by alternating the atomic displacements from layer to layer, and it is in the same flat phonon branch as $\Gamma_4^-$ [Fig.~\ref{f2} (a), (e), and (i)]. 
The structure obtained from activating the $Z_3^+$ and $Y_2^+$ modes thus gives a structure that looks like oIII in a single layer of conventional cells but with the polar atomic displacements alternating from layer to layer. 

Indeed, all other 2-period supercells shown in Fig.~\ref{f3} (f-l) can be viewed as layered unit-cell-scale domains of well-known phases (or their distorted version) and understood by similar constructions involving pairs of modes in the same flat phonon branch. 
We first consider the $Pbca$ structure shown in Fig.~\ref{f3} (f).
This structure, which was also discussed in Ref.~\cite{Aramberri23p95}, can be viewed as unit-cell-scale domain structure of the m phase with atomically-thin domain wall width.
As shown in Fig. S16 (a), in the m phase, the $Y_2^+$ and $Y_4^+$ modes are activated.
In the $Pbca$ structure shown in Fig.~\ref{f3} (f), the zone boundary mode $T_3^-$ is activated instead of the closely related zone center mode $Y_4^+$ in the same flat phonon band [Fig. S16 (b)].
The two m phases are related by a mirror reflection along the $z$ axis.
The $C2/c$ structure shown in Fig.~\ref{f3} (g) is another superlattice formed by two m phases, which are related by mirror reflections along both the $x$ and $z$ axes (See SM Fig. S17 for the mode analysis).

Fig.~\ref{f3} (h) is a superlattice with the $Pc$ space group, which is composed of the m and oIII phases with a sharp boundary. 
As shown in Fig. S18, the activated modes in the m/oIII superlattice includes the zone center modes $\Gamma_4^-$ and  $Y_4^+$ from the oIII phase [Fig. S18 (a)], as well as the $\Gamma_4^-$ and $Y_2^+$ modes from the m phase, along with two zone-boundary modes, $Z_3^+$ and $T_3^-$.
The combination of the $\Gamma_4^-$ mode [Fig.~\ref{f2} (e)] and $Z_3^+$ mode [Fig.~\ref{f2} (i)] leads to a polar cell (the oIII phase) and a non-polar cell (the m phase).
Similarly, the $Y_4^+$ [Fig.~\ref{f2} (b)] and $T_3^-$ [Fig.~\ref{f2} (f)] modes lead to the separation of oxygen atoms along the $y$ direction.
The combination of these two mode leads to a cell with a constructive interference and non-zero $y$-axis oxygen separation (the m phase) and a cell with a destructive interference and zero $y$-axis oxygen separation (the oIII phase). 
Since these two phonon bands are flat, the m and oIII phases can be connected with only a small distortion of each, forming a localized domain wall.

The $Cc$ structure in Fig.~\ref{f3} (i) is a superlattice formed by two oIII phases, with the right-hand one obtained by rotating 90$^{\circ}$ along the $y$ axis, followed by a mirror reflection across the $x$ axis, as shown in Fig. S23.
The $Pmna$ structure in Fig.~\ref{f3} (j) is a superlattice with an anti-polar structure which is formed by two oIV cells with opposite polarization. 
The formation of the last two supercells involves distorted monoclinic (denoted as dm) phases.
The $P2_1$ structure in Fig.~\ref{f3} (k) is a superlattice formed by connecting the oIII phase with a dm phase.
The relative distortion of the dm phase is illustrated in Fig. S24. 
First-principles calculations confirm that this phase is stabilized by the strain imposed by its neighboring cells (oIII phase here). 
It remains metastable when relaxed with fixed lattice vectors, but collapses into the m phase under full structural relaxation.
The $P2_1/c$ structure in Fig.~\ref{f3} (l) is formed by two dm phases, which are related by mirror reflections along all the three ($x$, $y$, and $z$) axes.
It has the same space group as the ground m phase, but has a higher energy (-1 meV/f.u. for this structure \textit{vs.} -80 meV/f.u. for the m phase).
All of these superlattice structures arise from activation of related modes from the same flat phonon branch, as discussed in detail in the SM section 8.

From this we see that HfO$_2$ has multiple competing polymorphs, which results from the multiple unstable phonon bands, taking the high symmetry $Cmma$ structure as the reference.
A specific polymorph forms due to the activation of one or several unstable phonon modes from flat branches. 
In the superlattices, the domain walls are atomically thin, a consequence of multiple unstable flat phonon bands. 
As shown in Fig. S26, such a flat unstable mode indicates that the high-symmetry $Cmma$ structure is unstable and equally favors relaxation into domain structures with arbitrary wavevectors. 
As a result, each cell can have an arbitrary mode amplitude, independent of its neighbors. 
Combining multiple unstable flat phonon modes, individual unit cell layers can adopt distinct phases independently, resulting in domains as narrow as a single unit cell with atomically thin domain walls.

\begin{figure}[hpbt]
\centering
\includegraphics[width=8.5cm]{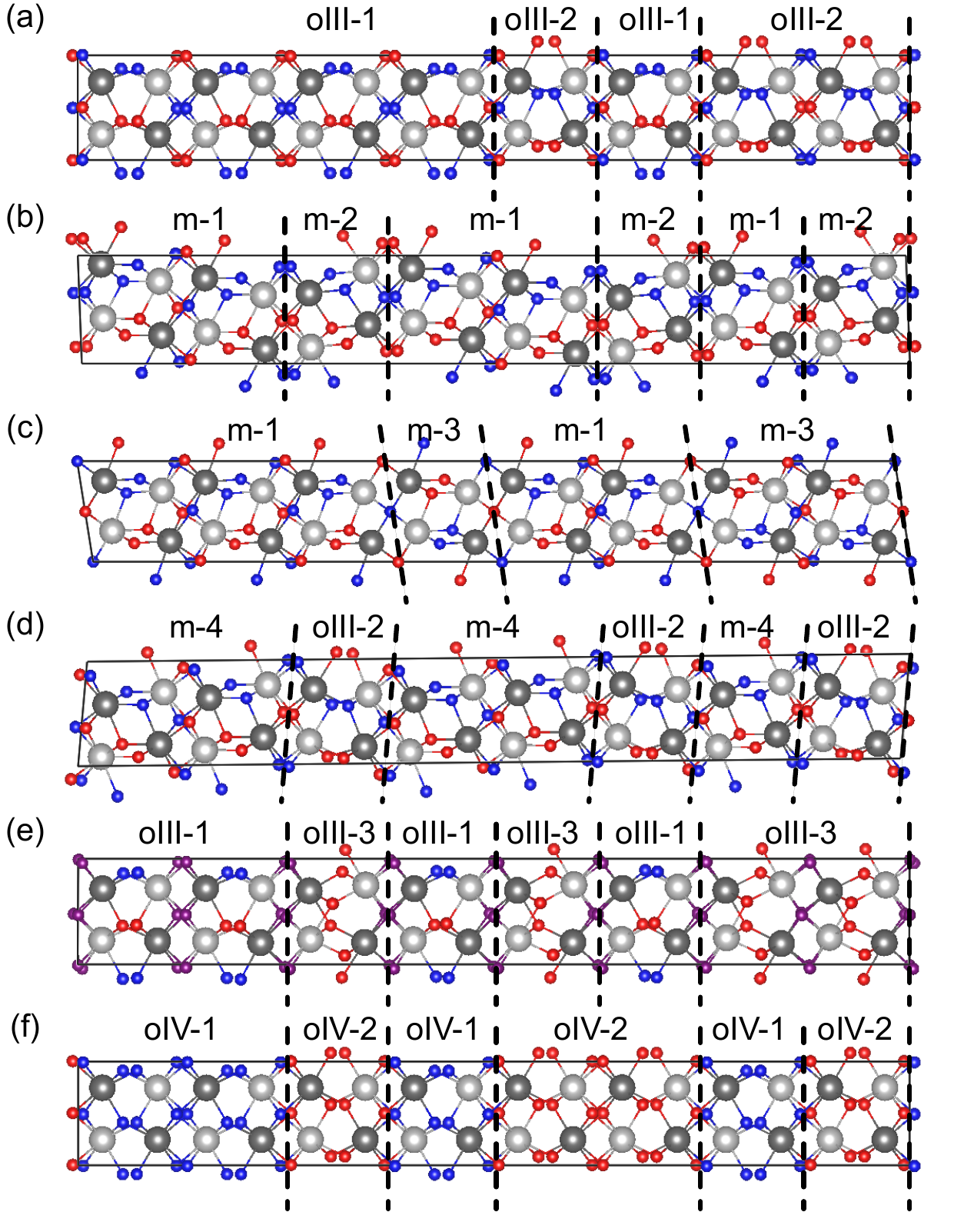}
\caption{(a-f) Fully relaxed superlattices with an arbitrary combinations of the two building blocks presented in the 2-period superlattice in Fig.~\ref{f3} (e-j).}
\label{f5}
\end{figure}

In our discussion above, we focused on the zone center and zone boundary modes, giving superlattices of period 2. 
In fact, the flatness of the phonon band from the zone center to the zone boundary allows the construction of superlattices with arbitrary period, which can be described as domain structures with atomically-thin domain walls.
Fully relaxed 2-period superlattices with various stacking orders of building blocks are shown in Fig.~\ref{f5}. 
Supercells containing dm phases are excluded, as a concentrated dm region combined with full relaxation may relieve the strain necessary to stabilize the dm phase.
These results explain why it is difficult to prepare a given phase of HfO$_2$ as a single crystal~\cite{Fina21p1530,Hsain22p010803,Xu21p826}.
Additionally, the locally stable superlattices with arbitrary periods and varied building blocks provide opportunities for engineering ferroelectric structures and their associated physical properties.

In summary, we demonstrate that the competing phases in  HfO$_2$ can be understood to arise from multiple unstable flat phonon bands in a high-symmetry $Cmma$ reference structure, identified from consideration of polarization reversal switching paths. 
Using systematic combination of the multiple unstable phonon modes and high-throughput first-principles calculations, we explored locally stable phases. 
This included not only the well-known phases, 
but new phases as well, which can largely be viewed as superlattices of the well-known phase. 
The unstable flat bands explain atomically thin grain boundaries in HfO$_2$ and also predicts the existence of superlattice phases of HfO$_2$ with arbitrary periods.
This work deepens the understanding of underlying physics in HfO$_2$ and provides guidance for functional phase stabilization.

\section*{acknowledgement}
Y. Q. acknowledges support from the National Science Foundation under Grant No. OIA-2428751
K.M.R. acknowledges support from the Office of Naval Research through N00014-21-1-2107.
First-principles calculations were performed using the computational resources provided by the Rutgers University Parallel Computing (RUPC) clusters, the High-Performance Computing Modernization Office of the Department of Defense, and the Cheaha Supercomputer at the University of Alabama at Birmingham.

\bibliography{cite}
\end{document}